\documentclass[aps,prl,showpacs,floatfix,twocolumn]{revtex4}

\usepackage{graphicx}
\usepackage{dcolumn}
\usepackage{bm}
\usepackage[]{natbib}
\begin{document}


\title{Probing the interior of neutron stars with gravitational
waves}

\author{L.K. Tsui}
\author{P.T. Leung\footnote{Email:
ptleung@phy.cuhk.edu.hk}}
\affiliation{%
Physics Department and Institute of Theoretical Physics, The
Chinese University of Hong Kong, Shatin, Hong Kong SAR, China.
}%

\date{\today}
\def\tomega{ \tilde{\omega} }
\def\tOmega{ \tilde{\Omega} }
\def\tr{ \tilde{r} }
\def\tx{ \tilde{r}_* }
\def\tV{ \tilde{V} }
\def\tpsi{ \tilde{\psi} }
\def\trho{ \tilde{\rho} }
\def\tP{ \tilde{P} }
\def\tR{ \tilde{R} }
\def\tX{ \tilde{R}_* }
\def\tm{ \tilde{m} }
\def\tphi{ \tilde{\phi} }
\def\tnu{ \tilde{\nu} }
\def\tlam{ \tilde{\lambda}}
\def\tepsilon { \tilde{\epsilon}}
\def\tJ { \tilde{J} }
\def\tU{ \tilde{U} }
\def\cc{{\cal C}}
\def\rr{ {\rm r} }
\def\ri{ {\rm i} }
\begin{abstract}
We show here how the internal structure of a neutron star can be
inferred from its gravitational wave spectrum. Under the premise
that the frequencies and damping rates of a few $w$-mode
oscillations are found, we apply  an inversion scheme to determine
its mass, radius and density distribution. In addition, accurate
equation of state of nuclear matter can also be determined.
\end{abstract}

\maketitle

\section{Introduction}
Despite that direct detection of gravitational waves from violent
astronomical events such as coalescing binary neutron stars (or
black holes) and supernovas has not yet been realized, it is
generally believed that it can be done in a few years.
Gravitational wave detectors of different designs, including
resonant antennas (e.g. EXPLOPER and NIBOE), LIGO and LISA
(currently under construction), complement the sensitivity of one
another and joint observations among them are likely to increase
confidence and accuracy in detection (see, e.g.
\citep{Hughes_03,LIGO_data} and references therein). In addition
to providing cogent support for general relativity, gravitational
wave detection indeed opens up a new channel for us to survey
stellar objects far away from our galaxy
\citep{Hughes_03,LIGO_data}.

\begin{figure}
\includegraphics[angle=270,width=7.5cm]{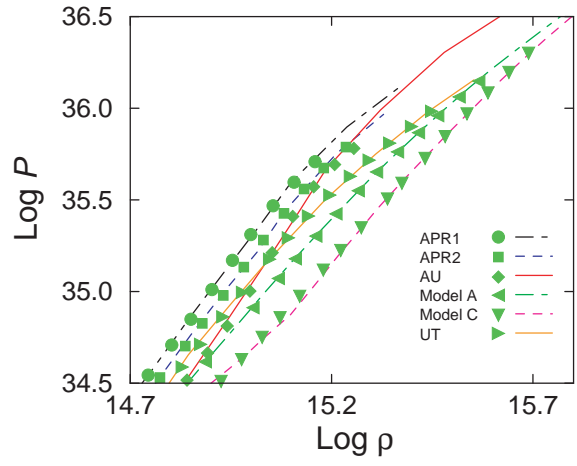}
\caption{Six different EOSs are shown by  lines of different
formats. Grey symbols are the corresponding values obtained from
inversion scheme using $\omega_1$ and $\omega_2$ for stars with
$\cc=0.28$. Here $P$ and $\rho$ are in cgs units.} \label{f1}
\end{figure}

Neutron stars and black holes are undoubtedly two major sources of
gravitational waves. However, unlike black holes that are
completely specified by a few physical quantities, the structure
of neutron stars is still masked because the behavior of nuclear
matter at high densities is not yet completely known. In fact,
there are currently numerous theoretical equations of state (EOSs)
of nuclear matter, leading to different
 models of neutron stars \citep{ComStar,Lattimer:2001,ss_crust,modelA,modelC,AU,APR}. To illustrate the
 differences in these EOSs, we show a few of them in
 Fig.~\ref{f1}, including models A and C
\cite{modelA,modelC}, models AU and UT \cite{AU}, models APR1 and
APR2 \cite{APR}.

As gravitational waves might carry the imprint of the internal
structure of a neutron star from which they are emitted, a close
examination of relevant signals, commonly analyzed in terms of
quasi-normal modes (QNMs) characterized by complex frequencies
$\omega=\omega_{\rm r}-\ri \omega_{\rm i}$
\citep{rmp,kokkotas_rev}, could possibly unveil its structure. In
fact, the term ``gravitational wave asteroseismology" has been
coined to indicate studies along this direction
\cite{Andersson1998,Ferrari}. Noting the universal behavior in the
QNMs of the fundamental fluid $f$-mode and the first polar
$w$-mode, Andersson and Kokkotas \cite{Andersson1998} showed that
the radius and the mass of a neutron star can be roughly estimated
from simultaneous detection of such wave modes. In addition, it
was argued that the frequency of the leading $p$-mode (or the
axial $w$-mode) oscillation could be used to identify an
appropriate EOS describing the neutron star
\cite{Andersson1998,Ferrari}.

 In a series of recent papers \citep{univ,pert}, we have
pointed out the physical origin of the universality in the
frequencies of the $f$-mode and the
 $w$-mode, and worked out
a scaled-coordinate logarithmic perturbative theory (SCLPT) to
locate axial $w$-mode oscillations. The main objective of this
Letter is to work out an inversion scheme to infer the internal
structure of a neutron star from its gravitational wave spectrum.
Under the premise that the frequency $\omega_{\rm r}$ and the
damping rate $\omega_{\rm i}$  of a few $w$-mode oscillations of a
neutron star can be identified from gravitational wave
observation, we iteratively invert SCLPT to determine the mass
$M$, the radius $R$ and the density distribution $\rho(r)$ of the
star. In particular, accurate EOS of nuclear matter can also be
obtained.
 While the idea underlying such inversion
scheme reported here is completely generic, we will consider and
use the spectrum of axial $w$-mode oscillations to illustrate our
method. Unless otherwise stated, geometric units are used in this
Letter.

\begin{figure}
\includegraphics[angle=270,width=7.5cm]{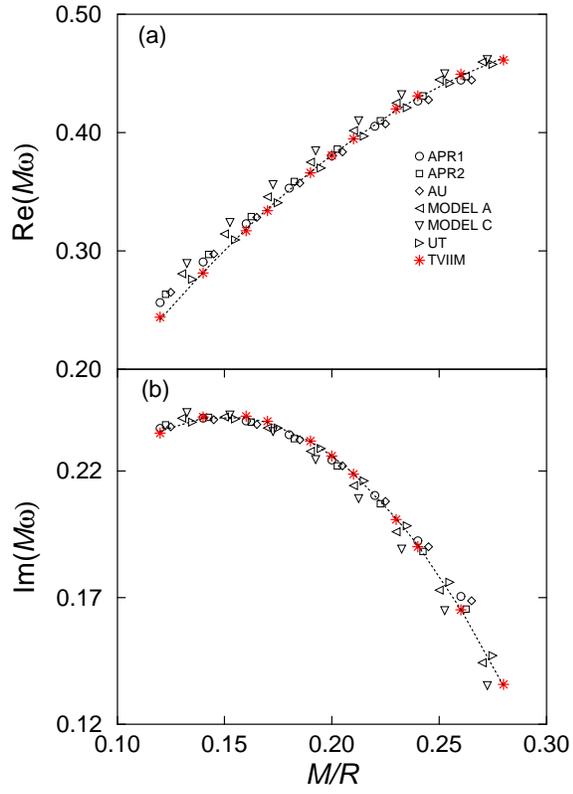}
\caption{The real and imaginary parts of $M\omega$ for the
least-damped axial $w$-mode of six realistic stars (unfilled
symbols) and TVIIM (stars) are shown as a function of $M/R$ in
panels (a) and (b) respectively. The dotted line represents the
best quadratic fit to those of TVIIM.} \label{f2}
\end{figure}

We first review the universality  in QNMs of neutron stars
\citep{univ}. As shown in Fig.~\ref{f2}, where $M\omega_{\rm r}$
and $M\omega_{\rm i}$ are plotted against the compactness $\cc
\equiv M/R$ for non-rotating neutron stars described by different
EOSs, the frequency of the $w$-mode (or $f$-mode) oscillation
approximately satisfies the following scaling law
\cite{Andersson1998,Ferrari,univ,pert}:
\begin{eqnarray}\label{TLA}
M \omega = a \cc^2+b\cc+c \,,
\end{eqnarray}
with $a$, $b$, and $c$ being complex constants determined from
curve fitting. Such universality  originates from the fact that
the mass distribution inside physical neutron stars can be
approximated by the Tolman VII model (TVIIM)
\citep{Tolman:1939jz}, whose mass distribution function
$m(r)\equiv 4\pi\int_0^r x^2\rho(x)dx$ is given by:
\begin{equation}
m_c(r)= M \left[ \frac{5}{2}\left(\frac{r}{R}\right)^3
-\frac{3}{2}\left(\frac{r}{R}\right)^5 \right].
\end{equation}
QNMs of TVIIM manifestly depend only on the compactness of the
star and reproduce the scaling behavior mentioned above
\cite{univ}. As shown in Fig.~\ref{f2}, TVIIM provides a good
approximation to stars with varying EOS and the universality in
(\ref{TLA}) can be captured by the best quadratic fit
 to the QNMs of TVIIM, with
  $a=-4.4-6.3{\rm
i}$, $b=3.1+1.9{\rm i}$, and $c=-0.072+0.098{\rm i} $
\citep{univ}.

As the first step of the inversion scheme, we study the frequency
of the leading (i.e. the least-damped) axial $w$-mode of TVIIM,
$\omega_1^{(c)}=\omega_{1\rr}^{(c)}-\ri\omega_{1\ri}^{(c)}$, and
find that the ratio $\omega_{1\rr}^{(c)}/\omega_{1\ri}^{(c)}$ is
in fact a monotonically increasing function of $M/R$ \citep{univ}.
Hence, once $\omega_1^{(c)}$ for TVIIM is known, $M/R$ can be
obtained from the ratio $\omega_{1\rr}^{(c)}/\omega_{1\ri}^{(c)}$
and in turn $M$ and $R$ can be found from (\ref{TLA}) or
Fig.~\ref{f2}.

As TVIIM indeed provides a benchmark for other realistic stars, we
expect that the frequency of the least-damped axial $w$-mode
emitted from a realistic neuron star, $\omega_1=\omega_{1\rr}-\ri
\omega_{1\ri}$, is close to that of the TVIIM star with the same
mass  and the same radius. Therefore, we could go through the
procedure outlined outlined above with $\omega_1^{(c)}$ replaced
by $\omega_1$ to obtain estimates of $M$,  $R$, $m(r)$ and hence
$\rho(r)$ for the star in consideration.

\begin{figure}
\includegraphics[angle=270,width=7.5cm]{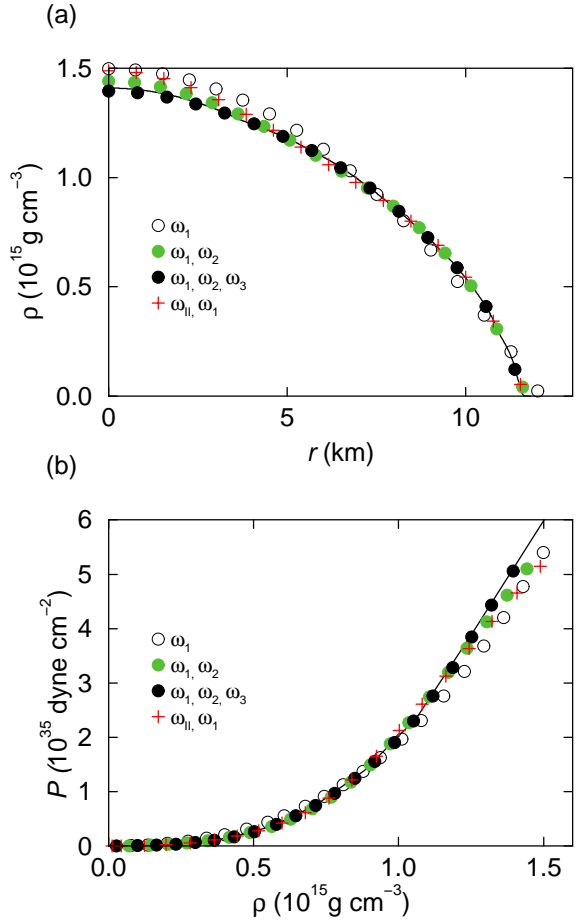}
\caption{Panels (a) and (b) depict $\rho(r)$ and $P(\rho)$
respectively for an APR1 star with $\cc=0.28$. The solid line is
the theoretical value, while the unfilled/grey/dark circles
represent the results obtained from inversion scheme using
one/two/three leading axial $w$-modes. The result obtained from
inversion scheme using $\omega_1$ and the frequency of a $w_{\rm
II}$-mode is shown by the crosses.} \label{f3}
\end{figure}

To gauge the accuracy of this scheme, we show in Fig.~\ref{f3}(a)
$\rho(r)$ of an APR1 star with $\cc=0.28$. The result obtained
from the above step (empty circles) agrees nicely with the exact
numerical data (the solid line). The pressure distribution $P(r)$
then follows directly from the Tolman-Oppenheimer-Volkoff (TOV)
equations \citep{Oppenheimer:1939ne,Tolman:1939jz} and in turn the
EOS $P(\rho)$ can be found. In Fig.~\ref{f3}(b) we show the result
of such scheme (empty circles), which satisfactorily reproduces
the EOS (solid line) except at high densities. In addition, as
TVIIM is exactly solvable \citep{Tolman:1939jz}, the EOS can be
found analytically:
\begin{eqnarray}
 P=\frac{1}{4\pi
R^2}\left[\tan\phi\sqrt{f(\cc)}
  -\cc\left(1+\frac{3\rho}{2\rho_0}\right)\right],
\end{eqnarray}
with $\rho_0\equiv \rho(r=0)=15\cc/(8\pi R^2)$,
\begin{equation}\label{}
f(\cc)\equiv3\cc -3{\cal C}^2
\left(1-{\rho}/{\rho_0}\right)\left(2+{3\rho}/{\rho_0}\right)\,,
\end{equation}
\begin{eqnarray}
\phi&=&\frac{1}{2}\log\left[\frac{1}{6}+
\sqrt{\frac{1-2\cc}{3\cal{C}}}\right]+
\tan^{-1}\sqrt{\frac{\cal{C}}{3(1-2\cal{C})}} \nonumber \\
&&- \frac{1}{2}\log\left[\frac{1}{6}-\frac{\rho}{\rho_0}+
\frac{\sqrt{f(\cc)}}{3\cc}\right].
\end{eqnarray}

It is worthwhile to note that a proposal has been put forward to
invert the EOS from given values of $M$ and $R$
\citep{Lindblom_invert}. In our scheme we make use of the values
of $\omega_{1\rr}$ and $\omega_{1\ri}$, which are considered as
data obtained from gravitational wave astronomy, to determine $M$
and $R$, and in turn obtain the EOS. Moreover, our scheme can
yield an analytic expression for the EOS while the proposal in
\citep{Lindblom_invert} has to resort to iterative numerical
solution.

In the following we will further consider the frequencies of
non-leading QNMs, $\omega_q=\omega_{q\rr}-{\ri} \omega_{q \ri}$
($q=2,3,\ldots$), ordered in increasing $\omega_{q \rr}$, and show
that they can lead to much improved estimates of the mass, the
radius, the density and EOS. As an incentive to the readers,  we
first display the accomplishment of our scheme. In Table~I we
tabulate the exact values of $M$ and $R$ and those obtained from
our scheme using one or two QNMs for stars constructed with
different EOSs. The result is truly encouraging, especially for
the mass. Hence, we conclude that the frequencies of two leading
$w$-mode oscillations can readily lead to accurate determination
of $M$ and $R$. In Fig.~1 we compare various EOSs and demonstrate
that it suffices to use only two frequencies ($\omega_1$ and
$\omega_2$) to accurately reproduce and distinguish these EOSs.
Figure~3 shows $\rho(r)$ and $P(\rho)$ obtained from the inversion
scheme using one, two and three QNMs, respectively. It is clearly
seen that the result obtained from a combination of $\omega_1$ and
$\omega_2$ (grey circles) are readily improved, while using all
three frequencies (dark circles) indeed yields a perfect match
with the exact values. We have also used $\omega_1$ and the
frequency of a $w_{\rm II}$-mode, $\omega_{\rm II}$, to infer the
physical quantities. However, the result is slightly less
accurate.

The technical details of our scheme are as follows. To improve on
TVIIM, we write $m(r)$ as:
\begin{equation}
m(r) \approx m_0(r;M,R)+\sum_{j=1}^{p-2}\mu_j m_j(r;M,R),
\end{equation}
where
\begin{equation}
m_j(r;M,R)=\frac{15Mr}{2R^5}\left(\frac{r}{R}\right)^{2j}(r^2-R^2)^2,
\end{equation}
$p\geq2$, and $\mu_j$ ($j=1,2,\ldots,p-2$) are adjustable
parameters to provide the best fit to $m(r)$. As $M$ and $R$ are
not yet exactly known, they are also considered as free
parameters. There are consequently $p$ free parameters to be
determined from the QNM frequencies. As mentioned above,
$m(r)\approx m_c(r)$ and therefore we will use $m_c(r)$ as an
initial guess for $m_0(r)$.

In the first step of our scheme, approximate values of the mass
and the radius of a star have been found from $\omega_1$, which
are equal to $M_0$ and $R_0$, respectively. Much improved values
of these quantities, $M=M_0+M_1$ and $R=R_0+R_1$, and the mass
distribution inside the star are now determined from higher-order
QNMs as follows. Suppose $n$ ($n>1$) QNM frequencies of a neutron
star are known. We look for optimal values of $\mu_j$
($j=1,2,\ldots,p-2$), $M_1\equiv \mu_{p-1} $, and $R_1\equiv
\mu_{p}$
  that yields frequencies with minimal deviations from
the measured ones. If all of these $p$ free parameters vanish, the
corresponding frequencies are $\omega_q^{(0)}$. Based on SCLPT,
the first-order change in $\omega_q$ ($q=1,2,\ldots,n$) is
\begin{equation}\label{cqj}
\omega_q^{(1)} =
\sum_{j=1}^{p}\mu_j\frac{\partial\omega_q}{\partial\mu_j}
=\sum_{j=1}^{p}\mu_j c_{qj},
\end{equation}
which is close to the exact change $
\Delta_{q}=\omega_q-\omega_{q}^{(0)}$. Explicit expression for
$c_{qj}$ can be found in \citep{pert}, which is expressible as a
bilinear form of the wave function.

By minimizing the total square-deviation $\chi^2$:
\begin{equation}
\chi^2
    \equiv \sum^{n}_{q=1}|\Delta_{q}-\omega_{q}^{(1)}|^2,
\end{equation}
we can obtain a set of $p$ linear equations for $\mu_{j}$:
\begin{equation}\label{matrix_A}
b_k = \sum_{j=1}^{p}A_{kj} \mu_j,
\end{equation}
where  $k=1,2,\ldots,p$ ,
\begin{eqnarray}
b_k &=& \sum^{n}_{q=1}\left( \Delta_q
    \bar{c}_{qk}
    +\bar{\Delta}_q c_{qk}\right), \\
A_{kj} &=& \sum^{n}_{q=1}\left( c_{qk}\bar{c}_{qj}
    + \bar{c}_{qk}c_{qj} \right).
\end{eqnarray}
and we have used an overbar to denote complex conjugation.
Consequently, $\{\mu_j\}$  can be computed.

Once $\{\mu_j\}$ are obtained, we can construct a new model star
by incorporating these perturbations into $m_0(r)$ and look for a
better approximation of the star by solving (\ref{matrix_A})
again. Through this iterative scheme,
 we can generate  rapidly convergent sequences for $M$, $R$, $\rho(r)$
 and $P(\rho)$.

\begin{figure}
\includegraphics[angle=270,width=7.5cm]{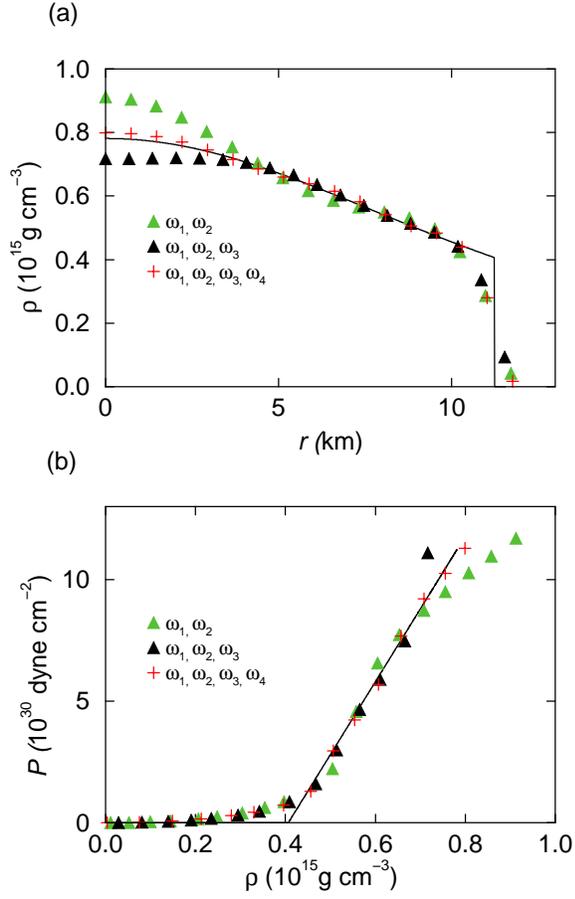}
\caption{Panels (a) and (b) depict $\rho(r)$ and $P(\rho)$
respectively for a strange star with $\cc=0.2$. The solid line is
the theoretical value, while the grey triangles/dark
triangles/crosses respectively represent the results obtained from
inversion scheme using two, three and four leading axial
$w$-modes.  } \label{f4}
\end{figure}

To demonstrate the robustness of the inversion scheme, we have
also applied our scheme to infer the EOS of a strange star
\citep{ComStar,ss_crust}. The EOS of quark matter,
$P=(\rho-4B)/3$, is given by the MIT bag model \citep{MITBM},
where $B= 57 \,{\rm MeV\, fm}^{-3}$ is the bag constant. The crust
of the strange star, where $\rho <$ neutron drip point, is
constructed with the Baym-Pethick-Sutherland EOS
\citep{ComStar,ss_crust}. The EOS for quark matter is rather
exotic and the density is even discontinuous across the inner
boundary of the crust. However, as shown in Fig.~\ref{f4}, the
inversion scheme still works nicely when four QNMs are used.

 Despite the mass of some neutron star binaries can be
inferred from their orbital periods, as yet there is no generic
method to determine their radii. In this Letter we have proposed a
robust inversion scheme to determine the mass, the radius, the
mass distribution and the EOS of a neutron star from its
gravitational wave spectra. We expect that our scheme could
operate in conjunction with gravitational wave detectors in the
near future to probe the interior of neutron stars.

Our work is supported in part by the Hong Kong Research Grants
Council (grant No: 401905) and a direct grant (Project ID:
2060260) from The Chinese University of Hong Kong.
\newcommand{\noopsort}[1]{} \newcommand{\printfirst}[2]{#1}
  \newcommand{\singleletter}[1]{#1} \newcommand{\switchargs}[2]{#2#1}

\newpage
\begin{table}
  \centering
  \begin{tabular}{|c|c|c|c|}
    \hline
    EOS &  $M$ ($10^{33}$ gm) & $R$ (km) \\
    \hline \hline
    APR1 & 4.460/4.475/4.460 & 11.83/12.12/11.70 \\
    \hline
    APR2 &  4.173/4.184/4.173 & 11.07/11.26/10.91 \\
    \hline
    AU & 3.843/3.862/3.843 & 10.19/10.61/10.08 \\
    \hline
    A & 3.289/3.282/3.289 & 8.723/8.529/8.564 \\
    \hline
    C & 2.791/2.788/2.791 & 7.404/7.185/7.216 \\
    \hline
    UT & 3.654/3.651/3.655 & 9.691/9.745/9.664 \\
             \hline
  \end{tabular}
  \caption{The mass $M$ and the radius $R$ of neutron stars constructed with
  different EOSs are compared with the corresponding values obtained from
  inversion using one or two QNMs. In each entry the values are
  obtained from TOV equation/inversion using $\omega_1$/inversion
  using $\omega_1$ and $\omega_2$ respectively.
  }
  \label{t1}
\end{table}
\clearpage
\newpage


\newpage 

\end{document}